# APPLYING ARTIFICIAL INTELLIGENCE AND INTERNET TECHNIQUES IN RURAL TOURISM DOMAIN

*Cristina TURCU, Cornel TURCU*

"Stefan cel Mare" University of Suceava
13, University Street, Suceava, 720225, Romania, tel.: +40-230-522978
e-mails: {cristina, cturcu}@eed.usv.ro

**Abstract**
    Society has become more dependent on automated intelligent systems; at the same time, these systems have become more and more complicated. Society's expectation regarding the capabilities and intelligence of such systems has also grown. We have become a more complicated society with more complicated problems. As the expectation of intelligent systems rises, we discover many more applications for artificial intelligence. Additionally, as the difficulty level and computational requirements of such problems rise, there is a need to distribute the problem solving. Although the field of multiagent systems (MAS) and distributed artificial intelligence (DAI) is relatively young, the importance and applicability of this technology for solving today's problems continue to grow. In multiagent systems the main goal is to provide fruitful cooperation among agents in order to enrich the support given to all user activities. This paper deals with the development of a multiagent system aimed at solving the reservation problems encountered in rural tourism. Due to their benefits over the last few years, online travel agencies have become a very useful instrument in planning vacations. A MAS concept (which is based on the Internet exploitation) can improve this activity and provide clients with a new, rapid and efficient way of making accommodation arrangements.

## Introduction

Tourism is sometimes believed to be a remedy for the development of rural areas [1]. Especially in the northern part of Romania where forestry and agriculture has lost their importance with respect to the number of people employed tourism based on natural resources, rural tourism is seen as the solution or part of the solution for creating jobs and a viable socio-economic situation.

Over the last few years, travel agencies have taken great pride in providing both individuals & companies with the highest quality service and cost effective savings when planning business or vacation travel arrangements. Constantly changing airfares and schedules, thousands of available vacation packages, and a vast amount of travel information on the Internet can make travel planning frustrating and time-consuming. To sort out the many travel options, tourists and business people often turn to travel agents, who assess their needs and help them make the best possible travel arrangements. In general, travel agents give advice on destinations and make arrangements for transportation, hotel accommodations, car rentals, tours, and recreation. For international travel, agents also provide information on customs regulations, required papers (passports, visas, etc), and currency exchange rates. Travel agents consult a variety of published and computer-based sources for information on departure and arrival times, fares, and hotel ratings and accommodations.

So a travel agency has become a full service organization that handles travel plans from start to finish. But in the Internet ages the personal computers can be used as private travel agencies. An online travel agency is more convenient than a traditional one because there are many advantages using it: many hotel searching and reservation services are on the web, which can offer online information about accommodation conditions and prices; also time schedules for air, sea, railway and ground transportation are available (all) over the world; no appointments and face to face meetings with travel agent are necessary. That's what a person gets when using the online services of the travel agency Web site. With a click of a mouse, one can access many travel agency benefits and services 24 hours a day, 365 days a year.

One important aspect in the rural tourism domain is related to booking. Traditional booking enforces the use of telephone or fax. Due to the several attempts to make a contact the first method is unsatisfactory. Also the fax method is characterized by large time delay. Using web based applications the booking process may be substantially improved. But the access to information is efficient when one knows where to find the information s/he is looking for. However, this task becomes complex when the information source is unknown. The complexity considerably increases when one does not know the type of information useful for his or her purpose and has only limited knowledge of the information sources put at his or her disposal without mentioning the fact that the arrangement and use of these sources can considerably vary from one to another. Thus, it becomes necessary either to ease the task of information retrieval, or to provide the users with some automation mechanisms [6].

## Multiagent systems

An agent is a software entity that applies Artificial Intelligence techniques to choose the best set of actions to perform in order to reach a goal specified by the user. It should react in a flexible, proactive, dynamic, autonomous and intelligent way to the changes produced in its environment. A multiagent system [8, 9] may be defined as a collection of autonomous agents that communicate between themselves to coordinate their activities in order to be able to solve collectively a problem that could not be tackled by any agent individually.

In the last years it has been argued that multiagent systems may be considered as the latest software engineering paradigm [2, 7]. This kind of systems may be used in domains with the following features [5]:
- Knowledge is distributed in different locations.
- Several entities, while keeping their autonomous behaviour, have to join their problem-solving abilities to be able to solve a complex problem.
- The problems in the domain may be decomposed in different sub-problems, even if they have some kind of inter-dependencies.

Due to the similarities with rural tourism domain the conclusion is that the multiagent systems seem to be adequate to be used in booking process.

## A multiagent system in rural tourism domain

During a booking process the user access a lot of distributed information sources that are now available on electronic supports. So it is necessary to create some tools to be used in retrieval information. To be efficient, such tools must be able to assess the context of an information request. The questions to be answered are the following: Who is the user? For what purpose is the research being undertaken? What are the relevance and quality of the retrieved information [3, 4]?

Obviously the implementation of an intelligent system can improve the whole booking process. Automated computer reservation systems may provide clients with instant access to thousands of guesthouse offers. Thus, clients can choose the best price and the fastest way of reaching their destination. Furthermore, one of the most important functions of this system would be the obtainment of best offers from guesthouses. Due to the rapid distribution of all travel information and the complexity of the planning process, intelligent systems can reduce the allocated time to a client and increase service quality.

In the last few years many guesthouses were built in the Romanian area. A lot of them are included into a national network named ANTREC[1]. Due to geographical distribution of the information one of the major problems occurres during the booking process. The development of an intelligent system requires the assessment of the agency's most important function: the client's request, alternatives to accommodation, notification on accommodation alternatives, reservation and payment. The temporal distribution of the reservation steps is presented in figure 1.

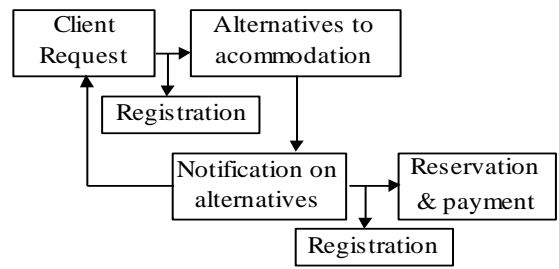

Figure 1. Reservation – temporal distribution

The system will be able to provide the following services:
- The user may ask for a reservation satisfying certain preferences (e.g. those guesthouses that are located in a particular area, price, facilities etc.).
- The user is able to access his/her historical record.

The proposed multiagent architecture is presented in figure 2. It consists in the following types of agents:
- *Personal agent*: Each *personal agent* represents a human user. The *personal agent* trustfully keeps information such as user profile and reservations made and is responsible for asking for possible bookings, in compliance with the users' preferences. It also controls the access of the *user interface agent* to the services provided by MAS, by avoiding direct communication with other agents.
- *User interface agent*: The *user interface agent* provides a graphical interface of the MAS with the user. This interface is used to introduce the requirements of a search or to show the results of a query to the user. It is necessary to define personalized interfaces that permit an easy, flexible and customizable access to the information the user need.

---

[1] ANTREC is a non-profit association that identifies, develops and promotes Romanian rural hospitality and tourism.

- *National agent*: it receives the users' preferences and sends a request to the *zonal agents*; also it receives all possible reservations and sends a classification to the *personal agent*.
- *Zonal agent*: For each zone there is one *zonal agent* and several *guesthouse agents*. The ZA has the general information of the guesthouses (e.g. its address, telephone number etc.).
- *Guesthouse agent:* It is one *guesthouse agent* for each guesthouse of a system. This agent could be executing in the desktop computer of each guesthouse. A *guesthouse agent* can encapsulate a web site, which provides reservation services (it access a database). It would also have a graphical interface that allows the personnel to update the information and request specific information from the database etc.

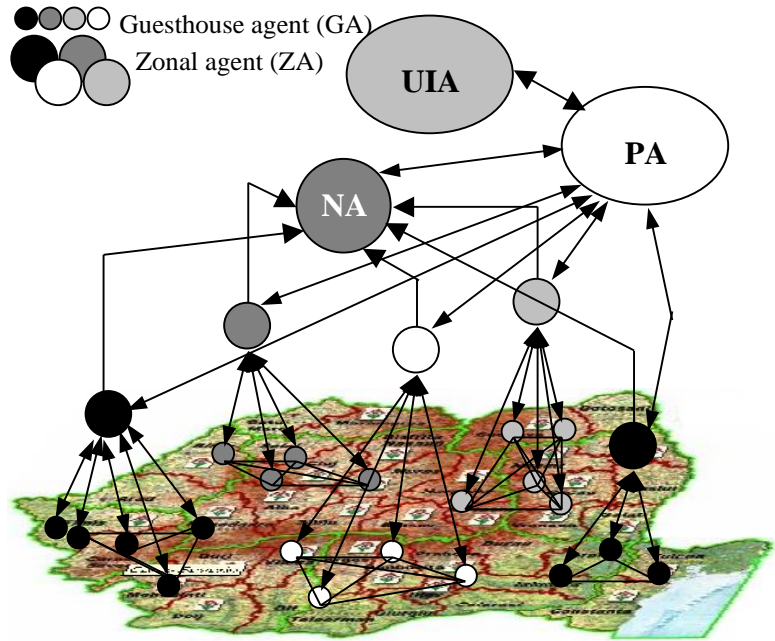

Figure 2. Multiagent system architecture (*PA*-Personal Agent, *UIA*-User Interface Agent, *NA*-National Agent)

Usually the system receives a user's request parameters set. These parameters refer to location wanted, number of persons, arrival & departure days, number of rooms wanted (single/double/triple), maximum price and other required facilities. If a guesthouse matches the user's preferences then the correspondent guesthouse agent will send an accepting message to *ZA*. The facilities of the guesthouse are presented in figure 3.

Figure 3. Guesthouse facilities

A possible reservation (from a zonal point of view) determination is presented in figure 4. In the represented situation the *zonal agent ZA1* receives the user's preferences and a request *id* (which is used as a parameter in all *ask* messages) from *national agent*. It will ask the *guesthouse agents* (*GA1_1, GA1_2,…*) for possible reservations. *GA1_3* does not match the user's preferences and send a *sorry* message to *ZA1*. The *GA1_2* agent does match the preferences and will send a *tell* message to *ZA1* about a possible reservation. The *GA1_1* agent does not match the preferences but it's possible to ask *GA1_2* and *GA1_3* for collaboration (for example, *GA1_1* cannot make a reservation for all the period requested by the user). *GA1_2* already made a reservation proposal (it checks *id* request) and will send a *sorry* message to *GA1_1*. If *GA1_3* can complete the period and match the user's preferences it will send a *tell* message to *ZA1*. So *ZA1* has two possible reservations: one from *GA1_2* and another one from *GA1_1* in collaboration with *GA1_3*.

The *NA* agent will receive all reservation proposals that meet the user's preferences from all *zonal agents*. After that it will reason about all of them and make a classification depending on one or more criteria (for example the price). The user will chose the best proposal and will initiate the booking process sending a message to the proper *GA* (through the MAS chain).

If the user specifies a requested zone, the entire reservation process will be made without *NA* participation. This implies the occurrence of direct communication between *PA* and *ZA* agents.

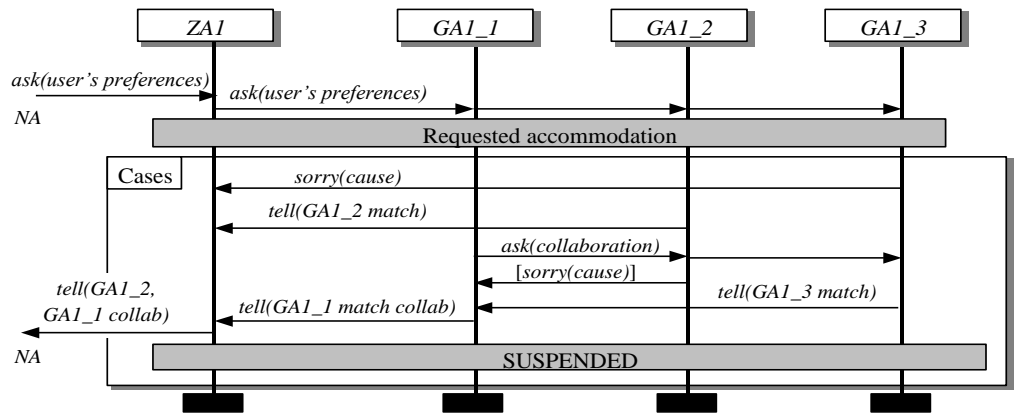

Figure 4. The reservation determination process

The system guarantees a secure access to information records, so that only authenticated users are allowed to update the information. There is a two-level security mechanism:
- Using a security plug-in of JADE (Java Agent Development Environment), JADE-S, all the messages between the agents in the system are encoded using SSL.
- The JADE-S module also allows the definition of different types of agents, and the assignment of different permissions to each kind of agent. For example, a *personal agent* cannot access directly any of the other agents in the system; it only has permissions to send and receive messages from *national agent* or *zonal agent*.

To ask information users may use diverse information technology. So, this system can be developed by considering the following easy and flexible communication ways: SMS messages through mobile phones, standard e-mail messages from a PC etc. In this case, it is necessary to add a new agent, named *Communication Manager Agent*. This agent is responsible of dealing with different types of information technologies. For example, the user might carry *Personal Digital Assistants* (PDAs) when they are on the way. We have also considered the possibility of attaching an intelligent agent to the PDA, who could request the information and send it wirelessly to the *communication manager agent*. The *communication manager agent*, after receiving the answer to the query could send this information to the agent permanently running in the PDA of a user.

## Conclusions

Due to the distribution of information the booking process in rural tourism domain is actually quite difficult. A solution is represented by a multiagent system approach. The multiagent system model seems to be the adequate framework for dealing with the design and development of an application, which is flexible, adaptable to the environment, versatile and robust enough to supply the booking process with efficiency and reliability.

The benefits of the system would reach two kinds of users:
- Citizens that need tourism information, who could obtain any kind of information related to the guesthouses of a given zone and make a reservation.
- Guesthouses personnel, who could access and update the guesthouse's information.

## References

[1]. Hall, C.M. & J. M. Jenkins, "The policy dimensions of rural tourism and recreation", In R. Butler, C.M. Hall and J. Jenkins (eds.) Tourism and Recreation in Rural Areas (pp. 19-42). Chichester: Wiley, 1998
[2]. Jennings, N., "On agent-based software engineering", Artificial Intelligence 117, 2000
[3]. King, J.A., "Intelligent Retrieval", AI Expert, 10.1, January, 15–17, 1995
[4]. King, J.A., "Intelligent Agents: Bringing Good Things to Life", AI Expert, February, 17–19, 1995
[5]. Moreno, A., "Agents applied in health care", Guest editorial, AI Commun 16(3):135-137, 2003
[6]. Pelletier, S., Pierre, S., Hoang, H., "Modeling a Multiagent System for Retrieving Information from Distributed Sources", Journal of Computing and Information Technology - CIT 11, 1, 1–10, 2003
[7]. Petrie, C., "Agent-based software engineering", in: Agent-Oriented Software Engineering, 2001
[8]. Weiss, G., "Multiagent systems. A modern approach to Distributed AI", M.I.T. Press, 1999
[9]. Wooldridge, M., "An introduction to Multi Agent Systems", Wiley Ed., 2002